\documentclass[sigconf,nonacm]{acmart}

\usepackage{enumitem}
\setlist{leftmargin=3.7mm}

\usepackage[T1]{fontenc}
\usepackage{algorithm2e}
\usepackage{graphicx}
\usepackage{textcomp}
\usepackage[capitalise]{cleveref}
\usepackage{enumitem}
\usepackage[frozencache,cachedir=.]{minted}
\usepackage{colortbl}
\usepackage{xcolor}
\usepackage{booktabs}
\usepackage[font=footnotesize]{caption}
\usepackage{multirow}

\definecolor{bggray}{rgb}{0.9,0.9,0.9}
\definecolor{bggreen}{rgb}{0.8,1.0,0.8}
\definecolor{bgred}{rgb}{1.0,0.8,0.8}

\newcommand{\code}[1]{{\small \texttt{#1}}}

\newcommand{\name}[0]{De-Hallu\-cina\-tor}

\setminted{frame=lines,breaklines,breaksymbol={\,},fontsize=\footnotesize,escapeinside=||,highlightcolor=bggray}
\RestyleAlgo{ruled}

\begin{document}
\lccode`\(`\(
\lccode`\)`\)
\lccode`\[`\[
\lccode`\]`\]
\title{\name{}: Mitigating LLM Hallucinations in Code Generation Tasks via Iterative Grounding}

\author{Aryaz Eghbali}
\affiliation{
    \department{Software Lab}
    \institution{University of Stuttgart}
    \city{Stuttgart}
    \country{Germany}
}
\email{aryaz.eghbali@iste.uni-stuttgart.de}
\author{Michael Pradel}
\affiliation{
    \department{Software Lab}
    \institution{University of Stuttgart}
    \city{Stuttgart}
    \country{Germany}
}
\email{michael@binaervarianz.de}

\begin{abstract}
    Large language models (LLMs) trained on datasets of publicly available source code have established a new state of the art in code generation tasks.
    However, these models are mostly unaware of the code that exists within a specific project, preventing the models from making good use of existing APIs.
    Instead, LLMs often invent, or ``hallucinate'', non-existent APIs or produce variants of already existing code.
    This paper presents \name{}, a technique that grounds the predictions of an LLM through a novel combination of retrieving suitable API references and iteratively querying the model with increasingly suitable context information in the prompt.
    The approach exploits the observation that predictions by LLMs often resemble the desired code, but they fail to correctly refer to already existing APIs.
    \name{} automatically identifies project-specific API references related to the model's initial predictions and adds these references into the prompt.
    Unlike retrieval-augmented generation (RAG), our approach uses the initial prediction(s) by the model to iteratively retrieve increasingly suitable API references.
    Our evaluation applies the approach to two tasks: predicting API usages in Python and generating tests in JavaScript.
    We show that \name{} consistently improves the generated code across five LLMs.
    In particular, the approach improves the edit distance by 23.3--50.6\% and the recall of correctly predicted API usages by 23.9--61.0\% for code completion, and improves the number of fixed tests that initially failed because of hallucinations by 63.2\%, resulting in a 15.5\% increase in statement coverage for test generation.
\end{abstract}

\maketitle

\section{Introduction}
\label{sec:intro}

Large language models (LLMs) have proven effective in many natural language~\cite{Brown2020} and programming tasks~\cite{Chen2021,codexStudy2022,DBLP:conf/sigsoft/XiaZ22,Poesia2022,Jain2022,Xu2022,Schaefer2024}.
Rapid adoption of LLM-based tools, such as Copilot\footnote{https://github.com/features/copilot} and Tabnine\footnote{https://www.tabnine.com/}, shows practical productivity benefits~\cite{Ziegler2024,Liang2024}.
State-of-the-art LLMs build on transformers~\cite{Vaswani2017}, which use self-attention to generate sequences of tokens in an auto-regressive process.
That is, the model decides what token to predict next based on the tokens in the prompt and any already generated tokens.
Hence, designing effective prompts, sometimes called prompt engineering, is a crucial part of developing a practical LLM-based technique~\cite{Liu2021a,Shrivastava2023,Nashid2023}.

Despite the impressive success of LLM-based code generation, these techniques are still at an early stage.
In particular, we identify two key challenges faced by current approaches:

\emph{Challenge 1: Project-specific APIs}.
As LLMs are trained on huge code bases, they effectively capture typical language idioms and commonly used libraries.
In contrast, a general-purpose model lacks knowledge of project-specific APIs, and may fail to correctly use existing functions and classes.
In particular, this lack of knowledge may cause the model to ``hallucinate'' APIs that actually do not exist in the current code base~\cite{Nguyen2022}, or perhaps even worse, it may reimplement some functionality that is already present in the code base.
Developers perceive this lack of knowledge about project-specific APIs as an obstacle to using AI programming assistants~\cite{Liang2024}.

As a running example, consider three files in a large project dealing with text documents, shown in \cref{fig:DataStore}.
One file, \code{DataStore.py}, contains a class implementing a data structure that stores documents and provides a keyword-based search over the documents.
Another file, \code{utils.py}, provides helper functions, one of which allows for measuring the relevance of a document to a keyword.
In a third file, \code{UI.py}, the developer is working on a function, \code{search}, to search for the \code{top\_k} documents that are most relevant to a keyword.

\begin{figure}[t]
    \centering
    \begin{minted}[label=DataStore.py]{python}
class DataStore():
  def __init__(self, file: str):
    with open(file, 'r') as f:
      self.documents = f.read().split('-----')
  ...
  def find_by_keyword(self, keyword: str) -> List[str]:
      return [d for d in self.documents if keyword in d]
  ...
\end{minted}
    \begin{minted}[label=utils.py]{python}
...
def relevance(document: str, keyword: str) -> float:
  return document.count(keyword) / len(document)
...
\end{minted}
    \begin{minted}[label=UI.py,highlightlines={4}]{python}
...
def search(ds: DataStore, keyword: str, top_k: int) -> List[str]:
  docs = ds.find_by_keyword(keyword)
  return sorted(docs, key=lambda d: relevance(d, keyword), reverse=True)[:top_k]
...
\end{minted}
    \caption{The desired completion of \code{search} is highlighted in \colorbox{gray!20}{gray}.}
    \label{fig:DataStore}
\end{figure}

\begin{figure}[t]
    \centering
    \begin{minted}[highlightlines={3}]{python}
def search(ds: DataStore, keyword: str, top_k: int) -> List[str]:
  docs = ds.find_by_keyword(keyword)
  return sorted(docs, key=lambda x: |\colorbox{red!20}{x.score\vphantom{,}}|, reverse=True)[:top_k]
\end{minted}
    \caption{The completion of \code{search} by CodeGen-2B-mono highlighted in \colorbox{gray!20}{gray}, and the wrong API usage highlighted in \colorbox{red!20}{red}.}
    \label{fig:codegen}
\end{figure}

Requesting an LLM, e.g., CodeGen~\cite{Nijkamp2022}, to complete the \code{search} function given a prompt that contains all existing code in \code{UI.py} results in \cref{fig:codegen}.
The code is partially correct, but refers to a non-existing API (an attribute \code{x.score}).
The underlying problem is that the models are not aware of the project-specific APIs that should be used to complete the code, and hence, the LLM simply hallucinates some plausible but ultimately wrong APIs.

\emph{Challenge 2: Prioritizing context}.
A naive solution to address Challenge~1 would be to simply add all of the code in the project into the prompt.
However, LLMs have a fixed maximum sequence length, which restricts how many tokens one can provide to the model.
Even with the recent increases in sequence length of LLMs, providing the most useful context can improve the output and reduce the costs.
Choosing the most helpful context for a given completion task is crucial, but an inherently difficult problem, because the optimal context depends on the desired code, which is not known a-priori.
While traditional code completion approaches typically have access to various kinds of information available in IDEs, such as the names and types of program elements, providing all this information, or even all the code of the project, to an LLM is impossible due to the limited prompt size.

This paper presents \name{}, which addresses the above challenges through a novel combination of retrieval-augmented generation (RAG) and an iterative form of LLM-based code generation.
Our approach uses three types of prompts, which provide increasingly suitable context information.
The \emph{initial prompt} type is querying the LLM with the conventional prompt, i.e., without any retrieval.
Retrieval-augmented generation (RAG)~\cite{Lewis2020rag} proposes to retrieve relevant context based on the initial prompt to address both Challenges 1 and 2, which we refer to as the \emph{RAG prompt} type.
The idea of augmenting an LLM with well-grounded facts relates to work on grounding of language models for natural languages~\cite{roy2005semiotic,ahn2022can,gu2022don}.
However, also this prompt may fail to generate factually correct code (i.e., without hallucinations), because the initial prompt might not have any similar code to the desired API, or there are other APIs more similar to the initial prompt than the correct one.
We make the improtant observation that the generated code from the previous prompt types often resembles the desired API.
Hence, we construct a new type of prompt called the \emph{iterative prompt}.
\name{} leverages the code that the model predicts to retrieve suitable project-specific APIs, which are then added to the iterative prompt for the next round of model prediction.
The iterative prompt type complements prior work that tries to guess the most suitable context from the incomplete code alone~\cite{Shrivastava2023,Ding2022}.

The presented approach offers several benefits.
First, \name{} works with any off-the-shelf LLM trained on code because the approach treats the model as a black box.
In particular, we do not require to train or fine-tune the model in any way, but simply exploit the fact that its predictions contain implicit hints about additional context the model would benefit from.
Second, because APIs usually evolve only slowly, \name{} can pre-compute, and occasionally update in the background, the set of project-specific API references.
As a result, the latency of code generation is not impacted by any expensive program analysis, which is important for practical adoption.
Finally, the approach is fully transparent to developers, because the approach hides the iterative interaction with the LLM from the user and simply returns a ranked list of predictions.

We evaluate \name{} by applying the approach to code completion with four state-of-the-art LLMs for code, namely CodeGen~\cite{Nijkamp2022}, CodeGen~2.5~\cite{Nijkamp2023}, UniXcoder~\cite{guo2022unixcoder}, and StarCoder+~\cite{li2023starcoder}, and to test generation with GPT-3.5-turbo.
Conceptually, the approach can be applied to any programming language, and our evaluation focuses on two popular languages, Python and JavaScript as they are among the most popular languages~\footnote{\url{https://octoverse.github.com/2022/top-programming-languages}} and common targets of prior work on code completion~\cite{DBLP:conf/ijcai/LiWLK18,svyatkovskiy2019pythia,Chen2021,guo2022unixcoder,Nijkamp2022,Nijkamp2023,li2023starcoder,Zhang2023b} and test generation~\cite{icse2022-Nessie,Schaefer2024}.
Compared to conventional prompts, we find that \name{} enables the models to provide more accurate predictions.
In particular, we show a relative improvement of 23.3--50.6\% in edit distance, and of 23.9--61.0\% in recall of correctly predicted API usages for code completion.
Moreover, we show relative improvement of 17.9\% in number of passing tests, of 15.5\% in coverage, and of 63.2\% in the number of mitigated hallucinations.

In summary, this paper contributes the following:
\begin{itemize}
    \item Empirical motivation showing that API hallucinations affect a large portion of failed code completion and test generation tasks.
    \item A technique for addressing this problem using off-the-shelf, unmodified LLMs.
    \item A novel algorithm that combines retrieval-augmented generation with an iterative method for constructing increasingly suitable prompts by using the hallucinations produced in earlier iterations to augment the context information provided in the prompts of future iterations.
    \item Empirical evidence that, across two code generation tasks, two programming languages, and five state-of-the-art LLMs, \name{} offers more accurate generations than conventional prompts.
\end{itemize}

\section{Preliminary Study}
\label{sec:preliminary study}

Before delving into our approach, we validate the motivation for this work by performing a preliminary study, which assesses the importance of the two challenges described in the introduction.

\subsection{Project-Specific APIs}

The main motivation for this work is our observation that LLMs often hallucinate code that resembles the desired code, but that fail to correctly refer to an API.
To assess the importance of this limitation, we investigate the prevalence and causes of hallucinated APIs in the two code generation tasks focused in this paper.
For code completion, we manually investigate and classify the reasons why an LLM fails to predict the desired completion.
We perform this preliminary study on 50 function-level code completion tasks, which we collect by (i) randomly selecting ten Python projects from a curated list of open-source projects~\footnote{\url{https://github.com/vinta/awesome-python}. We randomly sample ten application domains and then sample one project from each domain.} and (ii) by then randomly selecting five functions from each project.
The only filtering we perform is to ignore functions with more than 25 lines, as these are likely out of reach for today's LLMs.
For each of the 50 functions, we query an LLM (CodeGen 2.5 with 7B parameters and 4-bit quantization) with the code before the beginning of the function body, including the function signature and any docstring, in the prompt.

Given the 50 pairs of an LLM-predicted function body and the ground-truth function body, we manually classify them based on two questions.
First, is the prediction correct w.r.t.\ the ground truth, where ``correct'' includes exact matches and semantically equivalent code?
Second, does the ground truth contain an API usage, e.g., a function call, that is missing in the prediction?
Initially, two of the authors independently classify the 50 pairs, with an inter-rater agreement (Cohen's kappa) of 0.76, which is considered excellent~\cite{fleiss1981statistical}, and after discussing the discrepencies reach a consensus about all 50 pairs.

The final inspection results show that in 13 out of the 50 cases, the LLM either predicts exactly the expected function body or a function body that is semantically equivalent to the expected one.
For 22 out of the 37 remaining cases, there is at least one API usage that the LLM fails to correctly predict, similar to the example in \cref{fig:codegen}.
In other words, the problem identified and addressed in this work affects 44\% of all studied function-level code completion tasks, and even 59\% of all tasks where the LLM alone fails to predict the expected code.

For test generation, we use error messages of crashing tests generated by TestPilot~\cite{Schaefer2024} on a diverse set of 12 JavaScript projects.
We automatically count the number of generated tests that result in ``* is not a function'', or ``Cannot read properties of undefined'' errors, which typically indicates hallucinations of non-existing APIs.
We find that, on average, 16.4\% of all generated tests fail because of the aforementioned errors, which indicates that hallucinations of non-existing APIs are a common problem in test generation as well.

\subsection{Prioritizing Context}

To validate the importance of the second challenge, we compare the amount of code in a single project to the prompt sizes of high-end LLMs.
The models in the popular GPT series by OpenAI have prompt sizes between 4k (GPT-3.5 models) and 128k (GPT-4) tokens.
In contrast, in a sample dataset of 50 Python projects, which are randomly selected from the same curated list of projects as above, there are 488,635 tokens per project, on average.
Furthermore, the average project has around 13 files longer than 8,192 tokens, and 22 projects in our sample have at least one file longer than 32,768 tokens.
This means that even knowing the exact file that contains the relevant context (e.g., based on heuristics, such as recently used files or similar file names) leaves us with more tokens than one could fit into the prompt of some models.
Even for models with longer context window, considering a cost of 0.5\$ for 100k tokens means that the cost of long prompts would be impractical for regular use.
In other words, simply adding all potentially relevant code to the prompt is not a viable solution, but we need to prioritize the context information.

\section{Approach}

This section describes our approach for iteratively retrieving relevant APIs to improve the prompts for code generation tasks.
We call the approach \name{}, as it reduces the hallucinations of the LLM by providing relevant API references to ground the model.
First, we provide an overview of the approach (\cref{sec:overview}), and then present each of the components of \name{} in detail (\crefrange{sec:preprocess}{sec:llm}).

\subsection{Overview}
\label{sec:overview}

\begin{figure}[t]
    \centering
    \includegraphics[width=\linewidth]{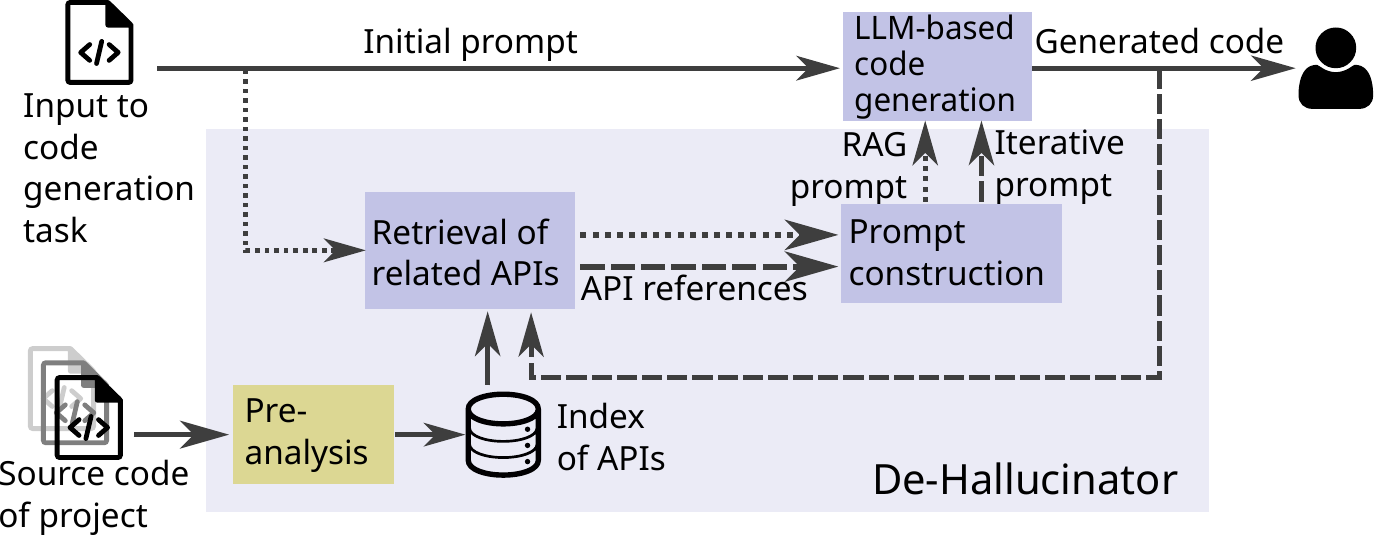}
    \caption{Overview of \name{}.}
    \label{fig:overview}
\end{figure}

\subsubsection{Main Algorithm}

\Cref{fig:overview} gives an overview of the approach, which we use to illustrate the main algorithm.
The top of the figure shows the traditional code generation process, where an LLM receives a prompt and generates code.
We call this prompt the \emph{initial prompt}.
Because the model may not be aware of project-specific APIs, the output is likely to refer to some hallucinated APIs.

To help the model predict better code, \name{} refines the initial prompt using two techniques.
Both of them add API references to the prompt, but they do so in different ways.
First, as shown by the dotted lines, \name{} uses the initial prompt to retrieve related API references from the project.
This approach is similar to retrieval-augmented generation (RAG)~\cite{Lewis2020rag}, and we refer to the resulting prompt as the \emph{RAG prompt}.

Second, as shown by the dashed lines, \name{} uses the output of the model to retrieve related API references.
This technique is unique to our approach, and it is based on the observation that the model's output often resembles the desired code, but fails to refer to the correct APIs.
Retrieving suitable API references based on the model's output can be done multiple times, and hence, we refer to the resulting prompt as the \emph{iterative prompt}.
In general, \name{} repeats the iterative prompt refinement, i.e., the dashed loop in the figure, until exhausting a configurable maximum number $k$ of queries to the model.

For efficient retrieval of APIs, \name{} analyzes the project in advance, as shown in the ``Pre-analysis'' component, and indexes all APIs of the project.

\subsubsection{Example}

\cref{fig:example_steps} shows each step of the approach on our running example from \cref{fig:DataStore}.
Given the initial prompt, the initial completion by the model refers to a non-existing API \code{x.score}.
Then, for the RAG prompt the initial prompt is used for retrieval.
In this step, as shown in \cref{fig:example_steps}, the reference to an already used function, \code{find\_by\_keyword} is retrieved.
Consequently, the completion uses the wrong API, as the model still does not have knowledge of the \code{relevance} function.
Next, using the initial completion by the model, \name{} retrieves a reference to the \code{relevance} function defined in \code{utils.py}.
In the example, the iterative prompt results in the correct completion, as shown at the bottom of \cref{fig:example_steps}.

\begin{figure}[t]
\centering
\begin{minted}[label=Initial prompt]{python}
...
def search(ds: DataStore, keyword: str, top_k: int) -> List[str]:
  docs = ds.find_by_keyword(keyword)
\end{minted}
\begin{minted}[label=Initial completion,highlightlines={4}]{python}
def search(ds: DataStore, keyword: str, top_k: int) -> List[str]:
  docs = ds.find_by_keyword(keyword)
  return sorted(docs, key=lambda x:|\colorbox{red!20}{x.score\vphantom{,}}|, reverse=True)[:top\_k]
\end{minted}
\begin{minted}[label=Most relevant API reference]{text}
DataStore.find_by_keyword(self, keyword: str) -> List[str]
\end{minted}
\begin{minted}[label=RAG prompt]{python}
# API Reference:
# DataStore.find_by_keyword(self, keyword: str) -> List[str]
def search(ds: DataStore, keyword: str, top_k: int) -> List[str]:
  docs = ds.find_by_keyword(keyword)    
\end{minted}
\begin{minted}[label=RAG prompt completion,highlightlines={4}]{python}
def search(ds: DataStore, keyword: str, top_k: int) -> List[str]:
  docs = ds.find_by_keyword(keyword)
  return sorted(docs, key=lambda x:|\colorbox{red!20}{x.score\vphantom{,}}|, reverse=True)[:top\_k]
\end{minted}
\begin{minted}[label=Most relevant API reference]{text}
relevance(document: str, keyword: str) -> float
\end{minted}
\begin{minted}[label=Iterative prompt]{python}
# API Reference:
# relevance(document: str, keyword: str) -> float
def search(ds: DataStore, keyword: str, top_k: int) -> List[str]:
  docs = ds.find_by_keyword(keyword)
\end{minted}
\begin{minted}[label=Iterative prompt completion,highlightcolor=bggreen,highlightlines={4}]{python}
def search(ds: DataStore, keyword: str, top_k: int) -> List[str]:
  docs = ds.find_by_keyword(keyword)
  return sorted(docs, key=lambda doc: relevance(keyword, doc), reverse=True)[:top\_k] # <- Correct
\end{minted}
    \caption{Step-by-step progression of \name{} on the example in \cref{fig:DataStore}.}
    \label{fig:example_steps}
\end{figure}

\medskip
\noindent
The following presents each component of \name{} in more detail, as well as how we instantiate the components for our two target tasks, code completion and test generation.

\subsection{Pre-Analysis}
\label{sec:preprocess}

\subsubsection{General Idea}
To ensure that the retrieval of API references does not unnecessarily slow down the code generation, \name{} has a preprocessing phase that indexes the current project for fast retrieval.
We use \emph{API reference} throughout this paper to refer to a piece of text extracted from the project's code, which can be added to the prompt to provide further information about a project-specific API.
\begin{definition}[API reference]
    An API reference is one of the following:
    \begin{itemize}
        \item A \emph{function reference}, which consists of
              \begin{itemize}
                  \item the qualified name of the function,
                  \item the parameter names,
                  \item any available default values for arguments,
                  \item any available type annotations, and
                  \item any available function-level docstring.
              \end{itemize}
        \item A \emph{class reference}, which consists of
              \begin{itemize}
                  \item the qualified name of the class,
                  \item the parent class(es), and
                  \item any available class-level docstring.
              \end{itemize}
        \item An \emph{attribute reference}, which is the qualified name of a \code{self} attribute assigned to in the constructor.
    \end{itemize}
\end{definition}

\subsubsection{Example}
\Cref{tab:APIref} shows some of the API references extracted from our example project.
Note that the API references resemble real code to maintain compatibility with any model that is trained on source code.

\begin{table}[t]
    \caption{Examples of API references extracted from the project in \cref{fig:DataStore}.}
    \label{tab:APIref}
    \centering
    \begin{tabular}{p{5em}p{13em}p{6em}}
        \toprule
        Source & API reference & Type of API reference \\
        \midrule
        \rowcolor{white}
        DataStore.py & \code{DataStore.find\_by\_keyword(self, keyword: str) -> List[str]} & Function reference    \\
        \rowcolor{gray!15}
        utils.py     & \code{relevance(document: str, keyword: str) -> float} & Function reference    \\
        \rowcolor{white}
        DataStore.py & \code{class DataStore()} & Class reference       \\
        \rowcolor{gray!15}
        DataStore.py & \code{DataStore.documents} & Attribute reference   \\
        \bottomrule
    \end{tabular}
\end{table}

\subsubsection{Application to Code Completion}
For the code completion task, we use CodeQL\footnote{https://codeql.github.com/} to statically extract APIs from the project source.
The Python language support of CodeQL offers easy access to the classes, functions, etc.\ in a code base.
Another benefit of CodeQL is that we can utilize databases created by GitHub for open-source projects.

\subsubsection{Application to Test Generation}
Since the APIs in JavaScript are in many cases only available once a module is instantiated, \name{} dynamically loads the modules and traverses them to extract API references.

\medskip
\noindent
Alternatively to our approaches for gathering API references, an IDE-based implementation could reuse information about the current project that is anyway computed by the static code indexing in an IDE.

\subsection{Retrieval of Related APIs}

\subsubsection{General Idea}
The retrieval module takes an input code piece and returns a ranked list of project-specific API references that are most similar to the input.
To enable similarity-based search, \name{} embeds the extracted API references into a vector space.
Formally, we need an embedding function, $E$, for which $E(c) = v_{c} \in \mathbb{R}^d$, such that, for two code pieces $c_1$ and $c_2$, the cosine similarity of their embeddings, $v_{c_1} \cdot v_{c_2} / (|v_{c_1}| |v_{c_2}|)$, approximates the semantic similarity of $c_1$ and $c_2$.
Recently, many models have been trained for this task~\cite{CodeEmbeddings}.
Because our approach uses the embedding function as a black-box, any embedding model or similarity-preserving vector representation~\cite{icse2021} can be used with \name{}, e.g., GloVe~\cite{Pennington2014}, BERT~\cite{Devlin2017}, or FastText~\cite{Bojanowski2017}.

Given the embeddings of the API references, \name{} retrieves API references that are most similar to the input code piece.
To this end, the approach embeds the input code piece and searches for the $n$ API references that minimize the cosine distance to the input code piece.
The parameter $n$ specifies the number of API references to include in the prompt.

\subsubsection{Example}

Getting back to our running example, consider the third section in \cref{fig:example_steps}.
It shows the API reference from our example project that the retrieval component finds to be the top-most relevant to the incomplete code (initial prompt).
Even though our implementation retrieves $n$ relevant API references, we show only one in the example for brevity.

\subsubsection{Application to Code Completion}
For code completion, we split the given input code piece into lines, then retrieve the most relevant APIs for each line, and finally merge the $n$ most similar API references into single sorted list.
The reason for retrieving the API references similar to full lines of code, as opposed to only API usages, is that we want the completion to avoid re-implementing existing code.
Therefore, if there exists some API similar to a line of code that does not use any APIs, we want the approach to be able to retrieve suitable API references to generate the correct completion.
We embed the API references into a vector space using Sentence-BERT~\cite{Reimers2019}, a BERT-based model designed for measuring the semantic similarity between sentences.
We use a variant of this model that is pre-trained on code.\footnote{https://huggingface.co/flax-sentence-embeddings/st-codesearch-distilroberta-base}
The model maps sentences, or in our case lines of code, into a dense vector representation of size 768.

After embedding the API references, the approach indexes the normalized vectors ($v_c / |v_c|$ for all $c \in \text{API references}$) in a Ball Tree.\footnote{https://scikit-learn.org/stable/modules/neighbors.html\#ball-tree}
This index allows for fast retrieval of nearest neighbors.
During the retrieval, we embed each line in the input using the same pre-trained SentenceBERT model used for indexing the API references, and then normalize the vectors.
The normalization is done to turn the Euclidean distance used by the Ball Tree into cosine similarity, which is commonly used.
Next, we find the closest API reference of each line by querying the Ball Tree constructed in the pre-analysis step.
The result is a list $R_l$ of API references, sorted by their similarity to the line $l$ in the input.
To obtain a single ranked list of API references, we merge the lists $R_l$ across all $l \in \mathit{completion}$ based on their similarity scores.
This finally yields a single list~$R$ of API references, of which we use the top-$n$ as additional context to add into the prompt.

\subsubsection{Application to Test Generation}
For test generation, we extract all API usages in a previously generated test using regular expression matching and embed them.
Next, we retrieve the most relevant APIs for every API usage and return a single list.
Here we retrieve API references based on API usages, and not based on lines.
The reason is that as there are no ground truths for the test generation task, and the main goal is to generate passing tests.
Therefore, fixing a wrong API usage is more important than avoiding re-implementation of existing code.
The embedding used for indexing the API references and during retrieval is a BERT-based model available on HuggingFace.\footnote{\url{https://huggingface.co/jinaai/jina-embeddings-v2-base-code}}
The reason for using a different model than the one used for code completion is that not all models on HuggingFace are compatible with TypeScript, so we choose a compatible model.
Using mean pooling, we embed the code pieces into a vector of size 768.
We use the same model for both indexing and retrieval, where during indexing each API signature is embedded and stored in a list, and during retrieval the API usage is embedded.
During retrieval, we perform a linear search for the most relevant API references.
Since the size of JavaScript projects are much smaller than the Python projects, there is not much benefit in using the Ball Tree data structure.

\subsection{Prompt Construction}
\label{sec:prompt}

\subsubsection{General Idea}
Given the input and a list of at most $n$ API references that may enable the LLM to accurately generate code, \name{} constructs an augmented prompt for querying the model.
The prompt is designed in a way that resembles ``normal'' code, i.e., the kind of data that the LLM has been trained on.
The API references are augmented as a block of commented lines to the prompt.
These lines start with \code{API Reference:}, and the following lines contain the relevant API references in decreasing order of similarity to the lines in the input of the retrieval module.

\subsubsection{Example}

For our running example, the ``Iterative prompt" section in \cref{fig:example_steps} shows the prompt for function \code{search} in our example, augmented with the API reference.
Given the iterative prompt, the same LLM that predicted the code in \cref{fig:codegen}, completes this function correctly in the last section of \cref{fig:example_steps}.

\subsubsection{Application to Code Completion}
In the code completion task, we prepend the prompt with the API references, as it minimally disrupts the structure of the existing code.
See our running example for illustration.

\subsubsection{Application to Test Generation}
On the other hand, for test generation, as TestPilot already includes additional information into the prompt, we append the API references to the end of the additional context section.
\Cref{fig:ex_test_gen} shows how the prompt is augmented with the API references in JavaScript, and the model's success in generating a correct test based on that prompt.

\subsection{Integration with the LLM}
\label{sec:llm}

\name{} is designed with minimal assumptions about the underlying LLM, and the tool that solves the code generation task.
The approach considers the LLM to be a black box that we query with a string, which then returns one or multiple strings with suggested code pieces.
Hence, we do not fine-tune the LLM, or train a model to preform retrievals, which makes \name{} applicable to more scenarios.
To that end, the instantiation of \name{} in both code completion and test generation tasks are compatible with any LLM, and our experiments with five LLMs show the flexibility of the approach.

\section{Implementation}
\label{sec:implementation}

The general ideas behind \name{} are language-agnostic, and the approach can be applied to different programming languages and to different code prediction tasks.
We present two implementations, one for code completion in Python and one for test generation in JavaScript.
The code completion implementation is in Python and builds on top of the HuggingFace transformers library.
Adapting our implementation to other models requires only to adjust the prompt size of the model and to select other parameters passed to its API.
The test generation implementation is in TypeScript and builds on top of the state-of-the-art LLM-based test generator TestPilot~\cite{Schaefer2024}.
Like TestPilot, we use GPT-3.5-turbo as the LLM.
TestPilot generates tests by going through the functions in the package under test, and for each function tries to generate a test with a simple input, which consists of a test header and the signature of the function under test.
Then, a set of ``prompt refiners'' modify the prompt to include usage snippets, error messages, and the body of the function under test.
We implement two new prompt refiners, one for the RAG prompts, and one for iterative prompts.
These two refiners are only activated when a test fails with an error that is likely caused by a hallucinated API, which we detect by checking if the error message contains ``is not a function'' or ``of undefined''.

\section{Evaluation}

To evaluate the effectiveness and efficiency of our approach, we perform experiments that answer the following research questions:
\begin{enumerate}[label=\textbf{RQ\arabic*:},leftmargin=8.6mm]
    \item How much does \name{} improve the generated code compared to the baseline approachs?
    \item How effective is \name{} at adding the correct API references to the prompt?
    \item How do the hyperparameters of \name{} affect the results?
    \item How efficient is \name{}, and how much do the different steps of the approach contribute to the running time?
\end{enumerate}

\subsection{Experimental Setup}

\subsubsection{Tasks}
Our evaluation targets two tasks, code completion and test generation.
For code completion, we define the task as completing an incomplete function at the beginning of a line with an API usage, given the preceding code and the existing code in the project.
This problem definition matches the common scenario of a developer implementing a function in an existing project, where the code to be written should use a project-specific API.
For example, suppose that the cursor in \cref{fig:DataStore} is at the beginning of the code marked with gray background. Everything above the cursor is our incomplete code~$c$, and the problem is to predict the marked code~$c'$, which refers, e.g., to the project-specific \code{relevance} API.
For test generation, the task is to generate tests for a given JavaScript package.
This problem setting has been well established by previous work~\cite{icse2022-Nessie,Schaefer2024}.

\subsubsection{LLMs and Baseline}

\paragraph{Code Completion}
We evaluate the code completion task on four state-of-the-art LLMs: CodeGen~\cite{Nijkamp2022} with 2.7B parameters (Sales\-force/code\-gen-2B-mono), CodeGen 2.5~\cite{Nijkamp2023} with 7B parameters (Sales\-force/code\-gen25-7b-mono), UniX\-Coder~\cite{guo2022unixcoder} with 125M parameters (micro\-soft/unix\-coder-base), and Star\-Coder+~\cite{Li2023} with 15.5B parameters (big\-code/star\-coder\-plus).
The reason for selecting these models is that they cover a variety of parameter sizes, model architectures, and pre-training processes.
We leave all parameters of the models at their defaults, except for the maximum new tokens parameter, which we set to 256 to allow for longer completions.
As a baseline, we query the models with a prompt that contains all the code preceding the cursor.
In case this prompt exceeds the maximum prompt size of 2,048 tokens, we truncate the prompt from the beginning.

\paragraph{Test Generation}
For the test generation task, we build upon TestPilot and use GPT-3.5-turbo-0125 for the LLM as the OpenAI GPT models are already integrated into TestPilot and require minimal effort to run.
As a baseline, we use the original TestPilot implementation with a one-hour time limit per package, 130k token limit per package, and four completions per prompt with temperature 0.1.
To make the comparison fair, we set the token limit of \name{} to the amount of tokens used by the baseline, and the number of completions to four with temperature 0.1. 
This prevents \name{} being advantaged with generating more tokens.
Moreover, we cache the model outputs so that the same prompts return the same completion for \name{} and the baseline.

\subsubsection{Datasets}
\label{sec:dataset}

\begin{table}[t]
    \caption{List of projects used for evaluation.}
    \label{tab:projects}
    \centering
    \begin{tabular}{llrr}
        \toprule
        Project (owner/name)          & Commit   & LoC     & Stars$^*$ \\
        \midrule
        \multicolumn{4}{l}{Python projects used for code completion}                                           \\
        \midrule
        \rowcolor{gray!10}
        graphql-python/graphene       & 57cbef6   & 9,484   & 7.8k      \\
        geopy/geopy                   & ef48a8c   & 10,000  & 4.3k      \\
        \rowcolor{gray!10}
        nvbn/thefuck                  & ceeaeab   & 12,181  & 83.3k     \\
        aaugustin/websockets$^{**}$   & ba1ed7a   & 14,186  & 5k        \\
        \rowcolor{gray!10}
        arrow-py/arrow                & 74a759b & 14,402  & 8.6k      \\
        lektor/lektor                 & be3c8cb & 16,852  & 3.8k      \\
        \rowcolor{gray!10}
        Parsely/streamparse           & aabd9d0   & 26,214  & 1.5k      \\
        Supervisor/supervisor         & ca54549 & 29,860  & 8.3k      \\
        \rowcolor{gray!10}
        mwaskom/seaborn               & f9827a3 & 37,367  & 12.1k     \\
        psf/black                     & ef6e079   & 106,005 & 37.6k     \\
        \rowcolor{gray!10}
        scikit-learn/scikit-learn     & f3c6fd6   & 193,863 & 58.5k     \\
        \midrule
        \multicolumn{4}{l}{JavaScript projects used for test generation}                                       \\
        \midrule
        \rowcolor{gray!10}
        node-red/node-red             & 29ed5b2   & 60      & 18.8k     \\
        winstonjs/winston             & c63a5ad   & 496     & 22.2k     \\
        \rowcolor{gray!10}
        prettier/prettier             & 7142cf3   & 916     & 48.5k     \\
        tj/commander.js               & 83c3f4e   & 1,134   & 26.3k     \\
        \rowcolor{gray!10}
        js-sdsl/js-sdsl               & 055866a   & 1,198   & 0.7k      \\
        goldfire/howler.js            & 003b917   & 1,319   & 23.1k     \\
        \rowcolor{gray!10}
        websockets/ws                 & b73b118   & 1,546   & 21.2k     \\
        handlebars-lang/handlebars.js & 8dc3d25   & 2,117   & 17.8k     \\
        \rowcolor{gray!10}
        petkaantonov/bluebird         & df70847   & 3,105   & 20.4k     \\
        hapijs/joi                    & 5b96852   & 4,149   & 20.7k     \\
        \rowcolor{gray!10}
        Unitech/pm2                   & a092db2   & 5,048   & 40.9k     \\
        11ty/eleventy                 & e71cb94   & 5,772   & 16.4k     \\
        \bottomrule
        \multicolumn{4}{l}{$^{*}$ As of June 5, 2024}                                                          \\
        \multicolumn{4}{l}{$^{**}$ Has been moved to python-websockets/websockets}                 \\
    \end{tabular}
\end{table}

\paragraph{Code Completion}
With the goal of having a diverse set of projects in terms of size, domain, and popularity,
we gather a dataset of eleven public Python projects from GitHub, shown in \cref{tab:projects} for the code completion task.
We construct a dataset of API-related code completion tasks by removing API usages from the benchmark projects and by considering the removed code as the ground truth to be predicted by a model.
For each such API call, we remove the lines containing the call.
If a call spans multiple lines, we remove all of them.
To prevent data leakage from imports of the API in the ground truth, we also remove API-related imports.
Next, we check if the off-the-shelf LLMs can predict the exact code as in the original file using the code preceding the cursor as the prompt.
If an LLM predicts exactly the original code, we ignore this API usage for the evaluation, as there is no need to further improve the prediction and to avoid any potential memorizations.
We continue with this process for each of the four models, until we have ten code completion tasks for each of the eleven projects.
During this process, we ignore 18, 51, 76, and 31 completions for UniXcoder, CodeGen, CodeGen v2.5, and StarCoder+, respectively.
Overall, the code completion evaluation dataset consists of 11 projects $\times$ 10 $\times$ 4 models = 440 code completion tasks.

\paragraph{Test Generation}
For the test generation task, we also gather a diverse set of 12 JavaScript projects from GitHub, shown in \cref{tab:projects}.
These projects cover a variety of domains, such as website generation, code formatting, and process management.
Since TestPilot cannot generate tests for ES modules, we only consider JavaScript projects that are CommonJS packages.
The rest of the setup is the same as in the TestPilot paper~\cite{Schaefer2024}.

\subsubsection{Metrics}

\paragraph{Code Completion}
We evaluate the code completions in three ways:
\begin{itemize}
    \item \emph{Edit distance}.
          To quantify the number of edits a developer needs to apply after receiving a code completion, we measure the edit distance between the predicted code and the ground truth.
          This metric provides a sense for how many token edits are saved when using \name{}.
          We compute edit distance using the Levenshtein distance at the subtoken level.
          For each pair of completion and ground truth, we tokenize the code pieces with a GPT-2 fast tokenizer,\footnote{https://huggingface.co/docs/transformers/model\_doc/gpt2\#transformers.GPT2TokenizerFast} and then calculate the edit distance using NLTK's \code{edit\_distance}.\footnote{https://www.nltk.org/\_modules/nltk/metrics/distance.html\#edit\_distance}

    \item \emph{Normalized edit similarity}.
          Similar to previous work~\cite{Lu2022} we also compute the normalized edit similarity.
          To this end, we normalize the absolute edit distance (computed as above) to the length of the longer of the two token sequences, and then turn the result into a similarity metric.

    \item \emph{Exact API match}.
          Since the goal of \name{} is to predict better API usages, we measure how many of all desired API usages are predicted exactly as in the ground truth.
          To identify the API usages in the lines of code to complete, we extract function calls, including the access path to the function, and the parameters.

\end{itemize}

For all the above metrics, we report the best completion obtained among $k$ completions of \name{}.
Measuring the $best@k$ matches a common usage scenario where a developer inspects a ranked list of code completion suggestions, and picks the first that matches the developer's expectations.

\paragraph{Test Generation}
For the test generation task, we use the following three metrics:
\begin{itemize}
    \item \emph{Number of passing tests}.
        We count the number of passing tests as a proxy for code without any hallucinations.
    \item \emph{Coverage}.
        We measure the statement coverage in the passing tests.
    \item \emph{Number of fixed hallucinated tests}.
        We measure the number of tests that initially crash with ``\textasteriskcentered{} is not a function'' or ``Reading property of undefined'' errors and that subsequently \name{} turns into passing tests.
\end{itemize}

\subsubsection{Hardware}

We perform the experiments with the CodeGen v2.5 model and the GPT-3.5-turbo-0125 model on a machine equipped with two Nvidia T4 GPUs, each having 16GB of memory.
The experiments with the UniXcoder, CodeGen, and the StarCoder+ models are performed on a machine with a single Nvidia Tesla V100 with 32GB of memory.
Each machine has a 48-core Intel Xeon CPU clocked at 2.20GHz.

\subsection{RQ1: Effectiveness of \name{}}
\label{sec:rq1}

\begin{table}[t]
    \caption{Effectiveness of \name{} in code completion compared to the baseline on four off-the-shelf LLMs. The \textbf{bold} numbers show statistically significant improvement over the initial prompt. The numbers in parentheses show the relative improvement over the baseline.}
    \label{tab:code_completion_results}
    \centering
    \begin{tabular}{@{}p{3em}p{4.8em}p{4.8em}p{4.8em}p{5em}@{}}
        \toprule
        Prompt type & UniXCoder 125M & CodeGen v1 2B & CodeGen v2.5 7B & StarCoder+ 15B \\
        \midrule
        \multicolumn{5}{@{}l@{}}{Edit distance (lower is better):} \\
        \midrule
        Initial & 52.4 & 40.0 & 47.2 & 44.6 \\
        RAG & \textbf{46.8} (10.7\%) & \textbf{33.4} (16.5\%) & \textbf{31.6} (33.0\%) & \textbf{35.9} (19.4\%) \\
        Iterative & \textbf{25.9} (50.6\%) & \textbf{30.7} (23.3\%) & \textbf{30.1} (36.3\%) & \textbf{33.5} (24.9\%) \\
        \midrule
        \multicolumn{5}{@{}l@{}}{Edit similarity (lower is better):} \\
        \midrule
        Initial & 33.4 & 43.6 & 43.9 & 33.2 \\
        RAG & \textbf{37.3} (11.7\%) & \textbf{48.0} (10.0\%) & \textbf{49.4} (12.5\%) & \textbf{38.0} (14.3\%) \\
        Iterative & \textbf{42.6} (27.5\%) & \textbf{48.9} (12.1\%) & \textbf{50.2} (14.2\%) & \textbf{39.7} (19.3\%) \\
        \midrule
        \multicolumn{5}{@{}l@{}}{Exact API match (higher is better):} \\
        \midrule
        Initial & 4.8 & 7.1 & 8.3 & 5.7 \\
        RAG & 4.8 (0.0\%) & \textbf{10.2} (42.6\%) & \textbf{11.6} (39.1\%) & \textbf{7.5} (32.0\%) \\
        Iterative & \textbf{5.9} (23.9\%) & \textbf{11.1} (55.3\%) & \textbf{13.4} (61.0\%) & \textbf{7.5} (32.0\%) \\
        \bottomrule
    \end{tabular}
\end{table}

\paragraph{Code Completion}
In the first set of experiments, we investigate to what extent \name{} improves code completions compared to the baseline.
By default, we run \name{} with $k=3$ iterations and add $n=20$ API references into the prompt.
As shown in \cref{tab:code_completion_results}, \name{} reduces the edit distance by 9.3 to 26.5 tokens, on average over the completions by the initial prompt, which is a relative improvement between 23.3\% and 50.6\% .
This in turn translates to normalized edit similarity improvements of 12.1\% to 27.5\% relative to the baseline.
Moreover, the approach relatively improves the exact API matches by 23.9\% to 61.0\%.
For example, for the CodeGen v2.5 model, \name{} is able to predict 1.5 times more APIs correctly than the baseline.
The approach shows statistically significant (using the Wilcoxon test and Pratt method) improvements over the baseline consistently for all metrics and all models.
For example, \Cref{fig:ex_dehallucinator_2} shows a scenario where \name{} improves the completion.
In this case the correct function is used by the model in the first try, but the order of parameters is wrong.
By providing the API reference in the prompt, \name{} predicts the correct API usage, as shown in \cref{fig:ex_dehallucinator_2}.

\begin{table}[t]
    \caption{Effectiveness of \name{} in test generation compared to TestPilot. The \textbf{bold} numbers show statistically significant improvement over the baseline. The numbers in parentheses show the relative improvement over the baseline.}
    \label{tab:test_generation_results}
    \centering
    \begin{tabular}{@{}p{8em}p{5em}p{5em}p{5em}@{}}
        \toprule
        Prompt type & Passing tests & Coverage & Fixed hallucinations \\
        \midrule
        Initial (TestPilot) & 64.8 & 32.1 & 19.3 \\
        RAG \& iterative & 66.3 (3.1\%) & \textbf{33.7} (3.6\%) & \textbf{43.2} (94.0\%) \\
        Iterative & 76.4 (17.9\%) & \textbf{37.0} (15.5\%) & \textbf{31.4} (63.2\%)\\
        \bottomrule
    \end{tabular}
\end{table}

\paragraph{Test Generation}
In the second set of experiments, we evaluate the effectiveness of \name{} in test generation.
We use $k=3$ and $n=3$ as default parameters for these experiments.
\Cref{tab:test_generation_results} shows statistically significant (using Wilcoxon test and Pratt method) improvements for code coverage and fixed hallucinations by \name{}.
Since RAG prompts use the initial prompt for retrieval, and because the initial prompt just contains a single signature, it is not as useful as iterative prompts.
This is also reflected in \cref{tab:test_generation_results} as lower coverage compared to only using iterative prompts.
Note that TestPilot uses some prompt refiners, such as retrying with error message, and including usage snippets and function bodies. These refiners can mitigate some hallucinations, but not as much as \name{}, as shown in \cref{sec:rq1}.
For example, \Cref{fig:ex_test_gen} shows a test generated by TestPilot, which uses non-existing APIs, but after providing the API reference in the prompt \name{} generates a test using the correct API.

\begin{figure}[t]
    \centering
    \begin{minted}{python}
async def schedule_formatting(sources: Set[Path], fast: bool, write_back: WriteBack, mode: Mode, report: "Report", loop: asyncio.AbstractEventLoop, executor: "Executor") -> None:
  """Run formatting of `sources` in parallel using the provided `executor`. (Use ProcessPoolExecutors for actual parallelism.) `write_back`, `fast`, and `mode` options are passed to :func:`format_file_in_place`.
  """
  cache: Cache = {}
  if write_back not in (WriteBack.DIFF, WriteBack.COLOR_DIFF):
    cache = read_cache(mode)
  |\colorbox{red!20}{sources, cached = filter\_cached(sources, cache)}| # <- baseline
  |\colorbox{green!20}{sources, cached = filter\_cached(cache, sources)}| # <- ground truth
\end{minted}
    \begin{minted}{python}
# API Reference:
# filter_cached(cache: Cache, sources: Iterable[Path]) -> Tuple[Set[Path], Set[Path]] # Split an iterable of paths in `sources` into two sets. The first contains paths of files that modifi
...
async def schedule_formatting(sources: Set[Path], fast: bool, write_back: WriteBack, mode: Mode, report: "Report", loop: asyncio.AbstractEventLoop, executor: "Executor") -> None:
  """Run formatting of `sources` in parallel using the provided `executor`. (Use ProcessPoolExecutors for actual parallelism.) `write_back`, `fast`, and `mode` options are passed to :func:`format_file_in_place`.
  """
  cache: Cache = {}
  if write_back not in (WriteBack.DIFF, WriteBack.COLOR_DIFF):
    cache = read_cache(mode)
  |\colorbox{blue!20}{sources, cached = filter\_cached(cache, sources)}| #<-|\name{}|
\end{minted}
    \caption{Completion by CodeGen highlighted in \colorbox{red!20}{red}, the ground truth, highlighted in \colorbox{green!20}{green}, and the completion by \name{} after augmenting the prompt with relevant APIs highlighted in \colorbox{blue!20}{blue}.}
    \label{fig:ex_dehallucinator_2}
\end{figure}

\begin{figure}[t]
\centering
\begin{minted}[highlightcolor=bgred,highlightlines={11,14,17,19}]{javascript}
let mocha = require('mocha');
let assert = require('assert');
let js_sdsl = require('js-sdsl');
// class js-sdsl.Deque(t = [], i = 1 << 12)
describe('test js_sdsl', function() {
    it('test js-sdsl.Deque', function(done) {
        let deque = new js_sdsl.Deque([1, 2, 3, 4, 5]);
        assert.equal(deque.size(), 5);
        assert.equal(deque.front(), 1);
        assert.equal(deque.back(), 5);
        deque.push_back(6);
        assert.equal(deque.size(), 6);
        assert.equal(deque.back(), 6);
        deque.push_front(0);
        assert.equal(deque.size(), 7);
        assert.equal(deque.front(), 0);
        assert.equal(deque.pop_back(), 6);
        assert.equal(deque.size(), 6);
        assert.equal(deque.pop_front(), 0);
        assert.equal(deque.size(), 5);
        done();
    })
})
\end{minted}
\begin{minted}[highlightcolor=bggreen,highlightlines={13}]{javascript}
let mocha = require('mocha');
let assert = require('assert');
let js_sdsl = require('js-sdsl');
// class js-sdsl.Deque(t = [], i = 1 << 12)
// API Reference:
// js-sdsl.Deque.prototype.pushFront(t)
// js-sdsl.Deque.prototype.front()
// js-sdsl.Queue.prototype.front()
describe('test js_sdsl', function() {
    it('test js-sdsl.Deque', function(done) {
        let deque = new js_sdsl.Deque([1, 2, 3]);
        assert.equal(deque.front(), 1);
        deque.pushFront(0);
        assert.equal(deque.front(), 0);
        done();
    })
})
\end{minted}
\caption{Test generated by TestPilot using GPT-3.5-turbo (top), and the iterative prompt by \name{} resulting in correct usage of APIs (bottom).}
\label{fig:ex_test_gen}
\end{figure}

\subsection{RQ2: Correct Retrieval of API References}

\paragraph{Code Completion}
To better understand the effectiveness of \name{}, we investigate how often the approach successfully augments the prompt with the correct API references.
Answering this question for all code completion tasks and all LLMs is difficult, because comparing the API references to the API usages is non-trivial due to different ways of importing APIs and passing arguments.
Instead, for code completion, we manually inspect a sample of 20 completion tasks per LLM and count the number of times an API used in the ground truth is successfully added to the prompt by \name{}.

\begin{table}[t]
    \caption{Number of APIs correctly augmented in the prompt.}
    \label{tab:api_in_prompt}
    \centering
    \begin{tabular}{@{}llp{3em}llll@{}}
        \toprule
        \multirow{3}{*}{Model} & \multirow{3}{*}{Tasks} & \multicolumn{5}{c}{API usages} \\
        \cline{3-7} & & Missing & \multicolumn{4}{c}{Expected API ref. added} \\
        \cline{4-7} & & /wrong & RAG & Iter. 1 & Iter. 2 & Iter. 3 \\
        \midrule
        UniXcoder & 20 & 18 & 4 & 5 & 5 & 5 \\
        CodeGen v1 & 20 & 17 & 5 & 6 & 6 & 6 \\
        CodeGen v2.5 & 20 & 15 & 5 & 5 & 6 & 6 \\
        StarCoder+ & 20 & 17 & 2 & 3 & 3 & 3 \\
        \bottomrule
    \end{tabular}
\end{table}

In our inspected samples for code completion, there are in total between 22 and 25 API usages (a completion can contain multiple API usages).
The new prompt generated by our approach contains the correct API between two and six times, as shown in \cref{tab:api_in_prompt}.
Moreover, for CodeGen and CodeGen v2.5 there are five completion tasks where the completion from the initial prompt either misses the APIs or uses them incorrectly, but in the completions from RAG or iterative prompts, the API reference section of the prompt contains the correct API. The same happens for UniXcoder in four tasks and for StarCoder+ in two tasks.
For cases where the approach fails to add the correct API reference into the prompt, the main reason is that the initial completion has low relevance w.r.t.\ the ground truth.

\paragraph{Test Generation}
Since for the test generation task there are no ground truths for the generated code pieces, manual inspection is infeasible.
Instead, we measure the number of passing tests and the number of non-crashing failing tests from an iterative prompt, as a proxy of success for the retrieval.
Note that an iterative prompt is only created when an initial prompt causes a crash with one of the specified errors in \cref{sec:implementation}.
We observe that from the 622 instances where an iterative prompt is generated, 16.7\% of iterative prompts result in a passing test, and 26.8\% of iterative prompts result in a test without hallucinations.
These results are also in line with the manual inspections of the code completion task above.

Overall, the results show that \name{} is able to successfully augment the prompt with the correct API references in many cases.

\subsection{RQ3: Impact of the Hyperparameters}

\begin{figure}[t]
    \centering
    \includegraphics[width=\linewidth]{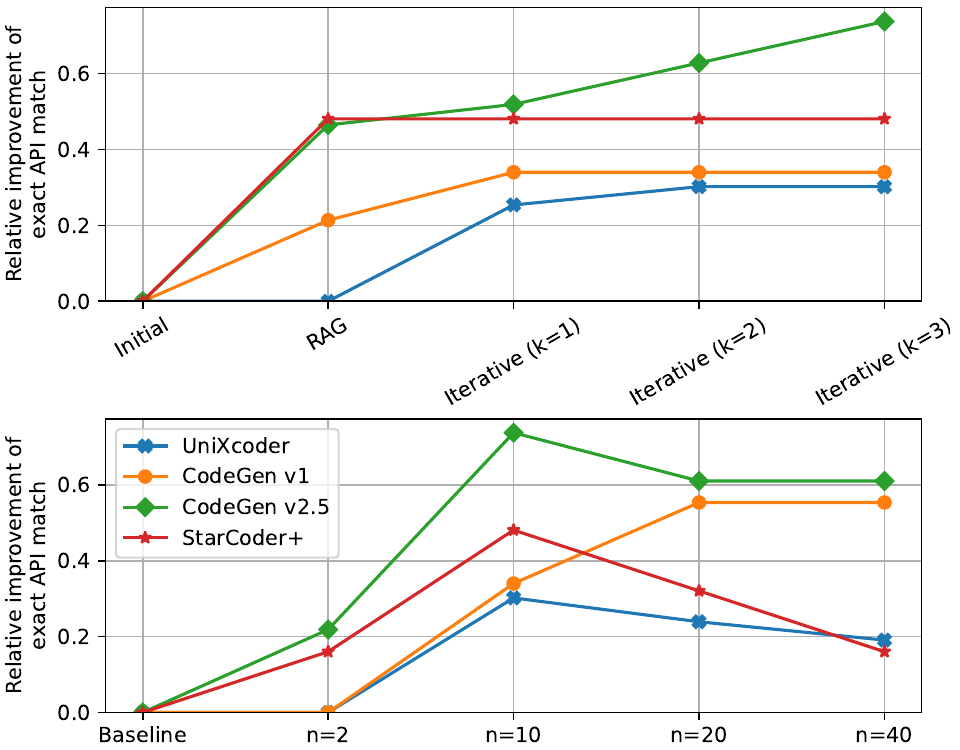}
    \caption{Effects of $k$ and $n$ for code completion.}
    \label{fig:effectkn}
\end{figure}

This research question evaluates the impact of the two main parameters of our approach.
First, we consider the number $k$ of iterations of iterative prompting.
For both code completion and test generation, we run the approach with $k \in \{1, 2, 3\}$.
As shown in \cref{fig:effectkn}, the first iteration provides significant improvement over the baseline, but the gain is reduced with further iterations.
Higher values of $k$ are beneficial when the model cannot immediately predict a relevant code, but upon presenting the first round of API references, the model responds with more relevant outputs.
At the same time, even $k=1$ provides clear improvements over the baseline, which makes \name{} useful even in scenarios where the cost of querying the model is high.
We choose $k=3$ for all other experiments in this paper.

\begin{figure}[t]
    \centering
    \includegraphics[width=\linewidth]{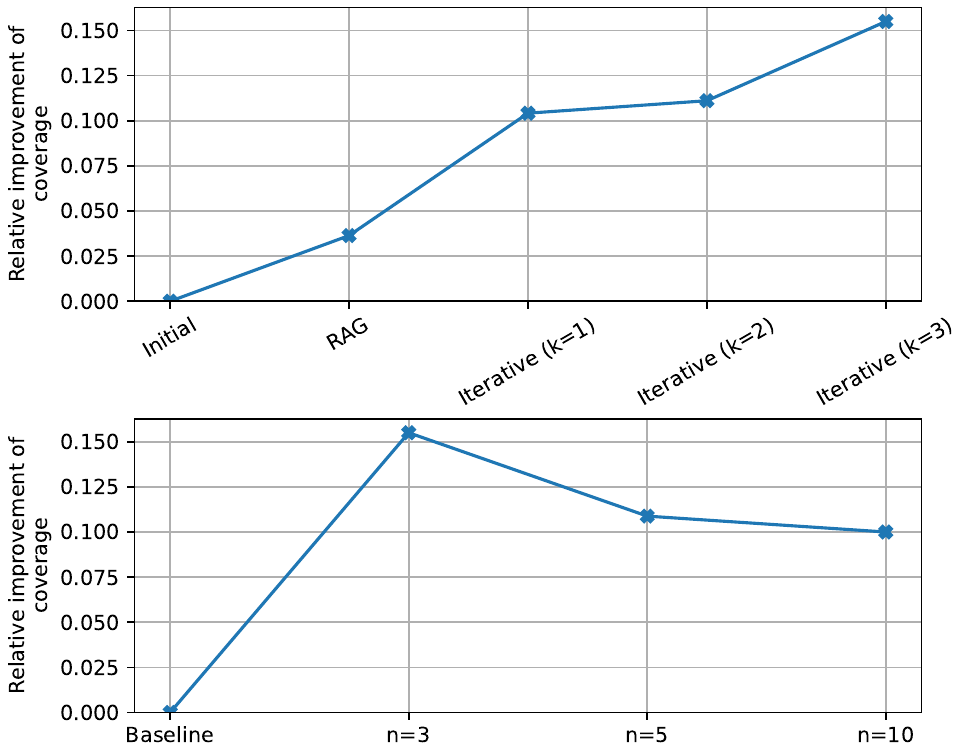}
    \caption{Effects of $k$ and $n$ for test generation.}
    \label{fig:effectkn_testgen}
\end{figure}

Second, we study the impact of the maximum number $n$ of API references that we add to the prompt.
Depending on the task, the optimal number of API references in the prompt varies.
Setting low values for $n$ can result in missing relevant context, whereas adding many API references can confuse the model, while also costing context space.
We perform experiments with $n \in \{2, 10, 20, 40\}$ for code completion, and $n \in \{3, 5, 10\}$ for test generation.
\Cref{fig:effectkn} shows the results of code completion, with exact API match peaking between $n=10$ and $n=20$.
As a default in the rest of the paper, we use $n=20$ for code completion.
\Cref{fig:effectkn_testgen} shows the effects of $n$ on coverage of the generated tests.
The peak in this case is at $n=3$, which we use as the parameter for other experiments in this paper.

\subsection{RQ4: Efficiency}
\label{sec:performance}

The following evaluates the efficiency of the approach and how much each of \name{}'s components contributes to its running time.
The pre-analysis step of code completion using CodeQL takes, on average, under one second per 1,000 lines of code in a project.
For the projects in our dataset, it takes at most 80 seconds, and most projects need at most 26 seconds for the whole preprocessing phase.
In a production-level implementation, our CodeQL-based approach could be replaced by using static information that is available in an IDE anyway, which is likely to further reduce the computational effort.
Moreover, updating the indexed API references, e.g., when the code base evolves, can be done at low frequency in the background.
For test generation, the pre-analysis takes between 0.6 seconds and 20 seconds with an average of 3.5 seconds.
Because the JavaScript projects in our dataset are smaller than the Python projects, the pre-analysis is faster for test generation.

Retrieving relevant APIs and constructing the augmented prompt for one iteration takes from 21 to 227 milliseconds for code completion, and from 0.1 to 17 milliseconds for test generation.
The time to query the LLMs ranges between 1.3 seconds (for the remotely deployed GPT-3.5) and 66.7 seconds (for CodeGen v2.5 running on our local Nvidia T4 GPU), on average per query.
These numbers are roughly the same for the baseline and for querying the model with \name{}-augmented prompts.
It is important to note that a production-level deployment would run the LLM on a GPU cluster, which typically answers queries within tens to hundreds of milliseconds, as evidenced by tools like Copilot and Tabnine.

\section{Limitations and Threats to Validity}

We assume an API to be available when generating the code that uses it.
However, in some cases, a developer may first write an API usage and then implement the API.
In such cases, \name{} would be unable to retrieve the API reference, and hence, could not provide any benefits.
To address this limitation, one could configure \name{} to abstain from repeatedly querying the LLM if the similarity between the initial completion and the retrieved API references is below a threshold.
We implement and evaluate \name{} for Python and JavaScript, and although our general approach could be applied to any language, our conclusions are valid only for these languages.
The set of projects we use might not be representative of all projects, which we try to mitigate by selecting a diverse set of popular projects.

\section{Related Work}

\paragraph{Data-Driven Code Completion}
The idea to augment traditional type-based code completion in a data-driven manner was introduced by Bruch et al.\cite{Bruch2009}.
More recently, statistical models are used, such as a pre-trained BERT model~\cite{Devlin2018} applied to code completion~\cite{Liu2020}, and models trained for specific kinds of completions, e.g., API usages~\cite{Raychev2014} and test methods~\cite{Nie2023}.
Grammars can improve statistical code completions, either by restricting the tokens to predict~\cite{Poesia2022} or by generating code that leaves some syntax subtrees undefined~\cite{Guo2022}.
Hellendoorn et al.~\cite{Hellendoorn2019} study data-driven code completion based on recorded real-world completion requests, with a focus on completing single identifiers.
Our work differs from all the above by providing project-specific API references based on previous completions as an input to a code completion model.

\paragraph{Code Completion with LLMs and Local Context}
Motivated by the observation that LLMs lack project-specific information, Shrivastava et al.~\cite{Shrivastava2023} propose a repository-level prompt generation technique to select the best context from a set of predefined contexts to solve the task of line completion.
Their method relies on training a separate model that takes a context window around the incomplete line as input, and outputs a ranking for additional contexts.
The training routine uses the LLM (in their case Codex) to calculate the loss function.
Ding et al.\cite{Ding2022} describe a similar method, called CoCoMIC, to address the challenge of project-specific APIs.
They utilize a custom static analyzer, CCFinder, that initially creates a context graph of program components in the project, and allows retrieval of relevant contexts to complete a statement.
They then fine-tuned CodeGen-2B-mono by adding the cross-file contexts to the input.
Both of the above are tightly coupled with the underlying LLM: The first approach~\cite{Shrivastava2023} uses the LLM to calculate the loss function for training a new model, and the second approach~\cite{Ding2022} changes the model's weights during fine-tuning.
In contrast, \name{} queries the LLM as a black-box, and hence, can be easily applied to other models.

An approach developed concurrently with ours~\cite{Zhang2023a} includes fragments of project-specific code in the prompt to improve the LLM's predictions.
Similar to our work, they also query the model iteratively.
Unlike \name{}, their approach retrieves existing code fragments, and not API signatures.
Since their approach relies on existing code fragments, it can only improve predictions when a project-specific API has already been used before and when this existing usage resembles the desired prediction, whereas our approach applies to all usages of project-specific APIs.

\paragraph{Combining LLMs and Retrieval}
Lu et al.\cite{Lu2022} use conventional retrieval methods to find similar code pieces in a pre-defined code database and add them to the prompt as dead code.
Although this approach improves the quality of code completions by the LLMs, it does not address the challenge of project-specific APIs.
Nashid et al.\cite{Nashid2023} propose a retrieval technique to find suitable examples for few-shot learning, but do not apply the idea to code completion.
HyDE~\cite{Gao2023} prompts an LLM to generate hypothetical textual documents for a given query, and then retrieves real documents that have an embedding similar to the hypothetical documents.
Their work shares the observation that LLM predictions may be factually inaccurate, e.g., in our case by referring to non-existing APIs, while being similar to a factually correct document.
By addressing this problem via retrieval, their approach is limited to producing already existing documents, whereas \name{} generates new code using an augmented prompt.

\paragraph{Automated Test Generation}
Before the era of LLMs, random feedback-directed test generation became practical through Randoop~\cite{Pacheco2007}.
LambdaTester~\cite{oopsla2018-LambdaTester} integrates higher order functions into test generation, and Nessie~\cite{icse2022-Nessie} targets asynchronous callbacks.
TestPilot~\cite{Schaefer2024} uses LLMs and prompt refinement to generate human-readable regression tests.
CodaMosa~\cite{Lemieux2023} uses LLMs to increase the coverage of automatic test generators when stuck in a plateau.

\paragraph{Improving LLM-Suggested Code}
To improve code suggested by LLMs, existing techniques for automated program repair~\cite{cacm2019-program-repair} can be applied in a post-processing step~\cite{Fan2022}.
Alternatively, the code predicted by a model can serve as input for initializing and guiding component-based code synthesis~\cite{Rahmani2021}.
The above work and ours shares the observation that completions from LLMs often share code elements with the desired code.
Instead of improving code in a post-processing step, \name{} nudges an LLM toward producing better completions by improving the prompt.

\paragraph{Querying LLMs Multiple Times}
Work on program repair queries a model multiple times until finding a suitable repair~\cite{Lutellier2020}.
They repeatedly query the model with the same prompt and may trigger thousands of queries, whereas \name{} continuously augments the prompt and queries the model only a few times.
Li et al.\cite{Li2022} propose querying a model with multiple mutations of the given code, and to then use the completion that is closest to the ``average'' completion.
\name{} instead uses the initial prediction to construct an improved prompt.
Xia et al.\cite{Xia2023} introduce conversational program repair, which iteratively improves a prompt by adding test failures observed when executing the predicted code.
In contrast, we do not require tests or executions, but only information that is statically available in a typical IDE.

\paragraph{Other Work on Models of Code}
The impressive abilities of neural models of code~\cite{NeuralSoftwareAnalysis} has lead to various other applications beyond code completion.
For example, neural models provide type predictions~\cite{Hellendoorn2018,icse2019,fse2020,Wei2020}, make predictions about code changes~\cite{Hoang2020,Brody2020}, and enable code search~\cite{Gu2018,Sachdev2018}.
LLMs are shown to be useful, e.g., for code mutation, test oracle generation, and test case generation~\cite{codexStudy2022,Schaefer2024,Kang2023}, and for automated program repair~\cite{Jiang2023}.

\section{Conclusion}

Motivated by the inability of current LLM-based code generation approaches to correctly predict project-specific APIs, we present \name{}.
The approach exploits the observation that LLMs often predict code that is similar to the desired code, but factually incorrect.
We address the hallucination problem by iteratively augmenting the prompt with increasingly relevant API references.
Our evaluation on two task, code completion and test generation, shows that \name{} significantly improves the quality of generations over the state-of-the-art baselines.

\section*{Data-Availability Statement}
Our implementation, datasets, and evaluation scripts are publicly available at \url{https://github.com/AryazE/dehallucinator} and \url{https://github.com/AryazE/testpilot}.

\bibliographystyle{ACM-Reference-Format}
\bibliography{references,referencesMP,referencesMore}

\end{document}